# Discovering mechanisms for materials microstructure optimization via reinforcement learning of a generative model

Rama K Vasudevan[1,*] Erick Orozco,[1,2,] and Sergei V. Kalinin[1]

[1] Center for Nanophase Materials Sciences, Oak Ridge National Laboratory, Oak Ridge, TN 37922

[2] Current affiliation: University of Southern California, Los Angeles, CA 90007

**Abstract**

The design of materials structure for optimizing functional properties and potentially, the discovery of novel behaviors is a keystone problem in materials science. In many cases microstructural models underpinning materials functionality are available and well understood. However, optimization of average properties via microstructural engineering often leads to combinatorically intractable problems. Here, we explore the use of the reinforcement learning (RL) for microstructure optimization targeting the discovery of the physical mechanisms behind enhanced functionalities. We illustrate that RL can provide insights into the mechanisms driving properties of interest in a 2D discrete Landau ferroelectrics simulator. Intriguingly, we find that non-trivial phenomena emerge if the rewards are assigned to favor physically impossible tasks, which we illustrate through rewarding RL agents to rotate polarization vectors to energetically unfavorable positions. We further find that strategies to induce polarization curl can be non-intuitive, based on analysis of learned agent policies. This study suggests that RL is a promising machine learning method for material design optimization tasks, and for better understanding the dynamics of microstructural simulations.

[*] vasudevanrk@ornl.gov

Properties of a broad variety of structural materials are ultimately underpinned by its microstructure, including phase morphologies, compositions, and interface properties. Among the historically recognized examples are unique mechanical properties of Damascus steel, one of the most closely-held secrets of Middle Ages. Another example of unique mechanical properties enabled via complex architectures are biological tissues, particularly bones.[1] Similar behaviors manifest for functional materials, including ferroelectric[2] and thermoelectric[3] materials, batteries and fuel cells,[4,5] and many others. For these materials, local morphologies control the phonon, electron, and chemical transport, enable mechanical stability and stress accommodation, and often give rise to novel interface-controlled functionalities. Critical for optimization of these materials is establishing the microstructure-property relationships and especially, the mechanisms behind the enhanced properties.

Similar challenges emerge on the nanometer and atomic level. Properties of spin and cluster glasses,[6,7] morphotropic ferroelectric materials,[8,9] and nanophase charge separated materials[10,11] are ultimately determined by microscopic details of the interactions between disorder and order parameters and electronic concentration, that give rise to degenerate, inhomogeneous ground states. The presence of almost degenerate states corresponding to strongly dissimilar ground states is one of the mechanisms behind enhanced properties, including giant magnetoresistance, strong electromechanical couplings, and dielectric responses. Similarly, the material design in these systems has come to the forefront of scientific research.[12]

Both for mesoscopic and atomically inhomogeneous systems, of interest are the details of local mesoscopic structure and mechanisms that connect microstructure to properties. Traditionally, local structures can be visualized using multiple imaging methods, ranging from optical and electron microscopy to X-Ray tomography and atom probe tomography. Similarly, local functionalities can be addressed by functional probes ranging from nanoindentation to tunneling spectroscopy or electron energy loss spectroscopy. Combining these data allows establishing local structure-property relationships, either via manual analysis or recently via machine learning methods. However, the limitation of the correlative approaches is that they are generally insufficient to address the interventional and counterfactual questions, e.g. predict the materials responses outside of the parameter interval at which measurements were taken, or predict the effects of the phase substitution or introducing the microstructures outside of the training set.

The alternative and far more established approach for exploration of these systems is physics-based analysis, aiming at the development of the lumped or distributed physical models that allow the construction of structure-property relationships. The example of the former are the classical mixture models for dielectric properties as a function of dielectric properties of individual phases. However, it is well recognized that many such descriptions are very sensitive to the specific morphologies, necessitating distributed physical models. For these, the classical example are the finite-element analyses for prediction of properties when the microstructure is known, and phase field models that allow to predict the morphogenesis and properties alike. However, while the distributed models allow to explore the arbitrary morphologies and resulting functionalities, the analysis of the response to derive relevant mechanisms is often intractable due to very high dimensionality of data (i.e. microstructure representation) and high computational complexity.

Here we propose an approach for the microstructure optimization towards target functionality and discovery of relevant physical mechanisms via reinforcement learning. We note that recently, RL has been utilized in design and optimization Kirigami structures[13], as well as for determining synthesis conditions in simulated environments.[14, 15] In this letter, we introduce the distributed model of the system and define the range of possible microstructure modifications. The reinforcement learning aims to learn the relationship between microstructure modifications and properties over multiple modification steps, thus precluding trapping in local minima. The resultant microstructures provide insight into the favorable physical mechanisms behind the property evolution.

**FerroSIM RL Environment Setup**

As a model system, we explore the lattice-based continuum model for ferroelectric proposed by Ricinschi et al.[16] and previously used for exploration of Bayesian optimization methods.[17] Noe that the underlying simulation used here has been described in reference.[17] Here we extend this model to create an RL environment for the training of agents for defect positioning. RL "environments" differ from conventional simulations in the sense that they allow external input to modify the trajectory of the simulation, and provide feedback in the form of rewards, as well as states to the agent. Briefly, the underlying ferroelectrics simulation is performed on a discrete pre-defined square lattice of side length N, with a free energy functional given by

$$F = \sum_{i,j}^{N} \left[ \left(\frac{\alpha_1}{2}\right) p_{x_{ij}}^2 + \left(\frac{\alpha_2}{4}\right) p_{x_{ij}}^4 + K \sum_{k,l} \left(p_{x_{ij}} - p_{x_{i+k,j+l}}\right)^2 + \left(\frac{\alpha_1}{2}\right) p_{y_{ij}}^2 + \left(\frac{\alpha_2}{4}\right) p_{y_{ij}}^4 + K \sum_{k,l} \left(p_{y_{ij}} - p_{y_{i+k,j+l}}\right)^2 - E_{loc_y} p_{y_{i,j}} - E_{loc_x} p_{x_{i,j}} \right] \quad (1)$$

Where $p_{x_{i,j}}, p_{y_{i,j}}$ are the *x*-component and *y*-component of polarization at the lattice site (*i,j*), *K* is a nearest-neighbor coupling term that effectively induces a gradient energy (with the sum taken over nearest neighbors of the lattice site, (*k,l*)), and the Landau parameters are $\alpha_1, \alpha_2$. Hence, this model effectively represents the Landau double well potential at each site, interacting via nearest-neighbor coupling. The local electric field $E_{loc} = E_{ext} + E_{dep} + E_d(i,j)$ where the first term is the externally applied electric field, the second term is a depolarization term $E_{dep} = -\alpha_{dep}\langle P \rangle$ where $\alpha_{dep}$ is a depolarization factor, and $\langle P \rangle$ is the average polarization, and the third term is the local random-field disorder induced by defects. As such, this simulator offers the possibility to explore both random bond disorder (through the coupling constant, *K*) as well as random field disorder effects.

Here, we focus on the effects of random field disorder. Each defect in the environment has the same *x,y* components of electric field (to within ~1%). Note also that here we consider *x* and *y* components to be decoupled entirely. The system is evolved based on the time-dependent Landau-Ginzburg equation, namely by computing the gradient at each lattice site via (2):

$$\frac{dp_{i,j}}{dt} = -\lambda^{-1}(\alpha_1 p_{i,j}^3 + \alpha_2 p_{i,j} + K \sum_{k,l}(p_{i,j} - p_{k,l}) - E_{loc}) \qquad (2)$$

Note that the dynamic equation (2) is calculated separately for the *x* and *y* components. This is then used to update the local polarization, $p_{i,j}(t_{n+1}) = p_{i,j}(t_n) + \Delta t \cdot \frac{dp_{i,j}}{dt}$ at each time step. The applied electric field is kept at 0 in this study, to study defect-induced polarization patterns. The simulation is sped up through only updating some fraction of lattice sites at every time step. This fraction can be changed, but we found that it was generally stable to only update 25% of the lattice sites, leading to about a 4x speedup. Future versions maybe able to incorporate order of magnitude greater speedups by leveraging GPUs and programming frameworks such as JAX. All simulation code is posted online.[18] Supplementary Table S1 lists the parameters of the simulation that are used in this paper. It should be noted that this is for all intents a "toy model" and the purpose is not to definitively perform simulated defect design on a standard ferroelectric but to show the potential for its use in real settings. That said, defect patterns discovered here could most certainly be used as inputs to more technically rigorous simulations, such as well-established phase-field models, when the necessary computational resources are available.

The RL environment is built on top of the simulation by providing intermittent rewards as well as states to the agent and allowing the agent to perform actions on the environment. The actions are to move any existing defect by one position on the lattice (up, right, left or down), as well as to 'pass' and not move any defects. This therefore defines a discrete action space, which is equal to *($n_d$ x 4 +1)*, where $n_d$ is the number of defects present in the simulation. The agent performs an action, and then the simulation runs from that point forward, returning a new state and (possibly) a reward. This continues until the end of the *episode*. The episode is terminated once a given number of moves have been made. This is pre-defined by the user.

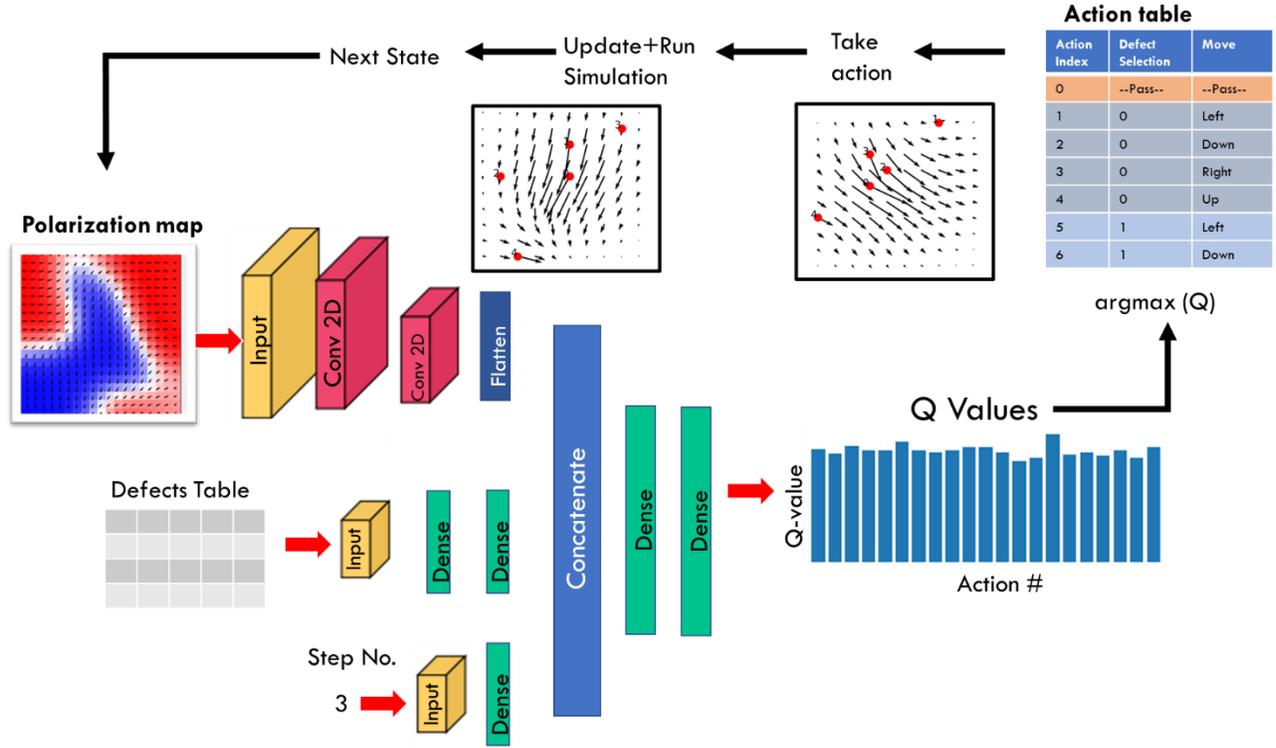

**Figure 1: Q-Network and State functions.** The polarization map, position of defects, and step number are all input to a neural network, which sends the input data through multiple convolutional and fully connected layers to generate a list of Q-values for that state. The index corresponding to the largest Q value is used and converted to an action (see Action table). The action is taken, and the simulation is updated and run for a pre-set number of time steps, to generate the next state. The process repeats until the episode ends, which is usually a pre-determined number of steps.

Several reward choices are built into the environment, including where the reward is the (absolute magnitude) of the curl of the polarization field, as well as the absolute magnitude of the remanent polarization, and one where the reward is the magnitude of polarization aligned to a specified orientation. In this environment, the reward can be given at the end of each episode or periodically (after each action) in the episode.

The general workflow for training RL agents in this environment is shown in **Fig. 1**. The state function itself consists of the 2D polarization map (which is a matrix of size $(2,N,N)$), with a tuple at each $(N,N)$ site denoting the $(P_x, P_y)$ components of polarization at that site, and a "defects table" which lists the position of the defects present, as well as their electric-field strengths. A third input is the step number in the episode. These are fed through a neural network with both convolutional and dense layers, and then outputted to estimate Q-values from which actions are selected. After an action is selected, this is used to update the simulation, and the simulation runs a certain number of time steps (typically 100-200) before halting to provide the next state and (if appropriate), reward.

**Deep Q Learning**

For the agents, we utilize deep Q learning to learn an implicit *policy* to select actions that maximize the cumulative discounted reward. *Policies* govern how agents act in the environment. Deep Q learning is a model-free method that is, compared to other RL methods, generally sample efficient. The idea behind Q learning is to learn the action-value function (or 'Q' function) that determines the value of taking a particular action at a given state. The Q function of policy $\pi$ is defined as

$$Q^\pi(s_t, a_t) = \mathbb{E}[r_t + \gamma r_{t+1} + \gamma^2 r_{t+2} + \cdots | s_t, a_t] \tag{3}$$

And computes the expected return of taking a given action, assuming the policy $\pi$ is followed. Note that $0 \leq \gamma \leq 1$ is a discount factor, which weights the value of immediate and future rewards, and the $Q$ function is an expectation conditioned on both the state and action. The simplest policy is one where the agent selects the action that provides the largest $Q$ value at every state. Note that we can rewrite (3) as:

$$Q^\pi(s_t, a_t) = r_t + \gamma \max_a Q(s_{t+1}, a) \tag{4}$$

After repeated interactions with the environment, we use the accumulated data on the states, actions, and rewards to update the $Q$ function via the simple weighted update rule,

$$Q'(s_t, a_t) \leftarrow Q(s_t, a_t) + \alpha_{lr} \cdot \left(r_t + \gamma \max_a Q(s_{t+1}, a) - Q(s_t, a_t)\right)$$

where $\alpha_{lr}$ is the learning rate. The second portion constitutes the loss that is minimized, and after many iterations, ideally the $Q$ function approaches the true action-value function. Deep $Q$ learning builds on $Q$-learning via several updates, including using neural networks as function approximators that can model this $Q$ function, and using so-called "experience replay" to reduce correlations in the data. This involves storing the "experiences", i.e., the states visited and actions taken and rewards received, of the agent in a memory bank and then sampling this experience in batches randomly for the $Q$-function update step. Here, we utilized the Adam optimizer for optimizing the neural network representing the $Q$ function ("$Q$-network"), and used experience replay with a batch size of 16-32, with a memory of 128-256 (problem dependent).

**Three Defect Curl Maximization**

As a test problem, we explore the best strategy of positioning defects to maximize polarization curl, which for a 2D situation is defined as $\nabla \times P = \frac{\partial P_y}{\partial x} - \frac{\partial P_x}{\partial y}$, resulting in a scalar field along the $z$-direction. The reward is given at the end of each episode, as the sum of the (absolute value) of this scalar field, divided by the magnitude of the polarization. Each episode starts with the polarization in a random configuration at each lattice site, and with three defects placed in locations shown in **Fig. 2**(a), i.e., the starting position of the defects does not change each episode. The agent has 15 moves in total each episode, and the simulation is run for 100 time steps after each move. The goal is to find the best way to behave on average, to accumulate the maximum reward. The stochasticity of the simulation means that this is a difficult optimization problem: small changes in initial polarization can make considerable changes to later time steps. Degeneracy of states is also a complicating factor, and the smaller number of iterations (100 as opposed to say, 1000) is insufficient to ensure that the actual minima is reached. That said, this scenario is highly reminiscent of realistic applications where the systems are often in kinetically-frozen states. For example, slow motion of charged defects in ferroelectric oxides has been instigated as a reason for "persistent" domain wall conductivity in $BiFeO_3$.[19]

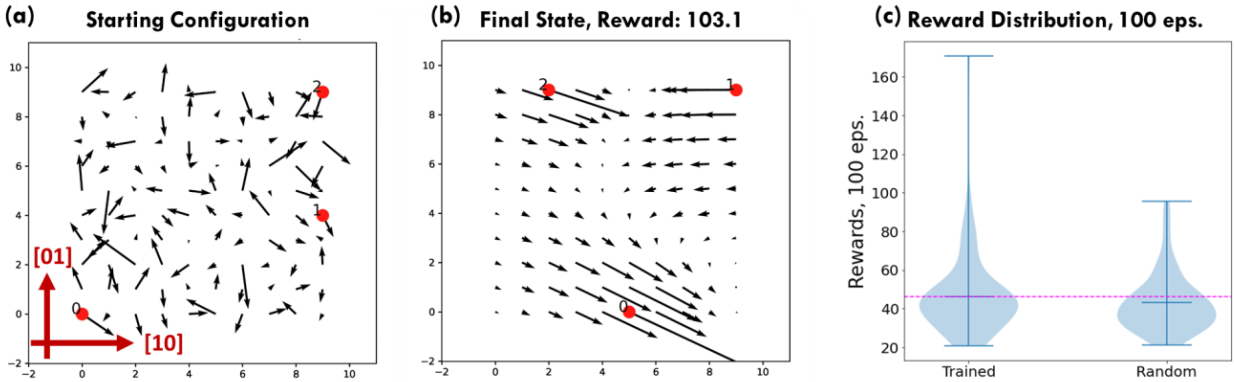

**Figure 2:** Maximizing curl via movement of three defects. (a) Starting configuration, before any time steps, shows the initial positions of the defects and the random initialization of the polarization at each lattice site. (b) Final state of one run of the trained agent, showing that the agent has preferred to move the defects to large triangular configuaration. (c) Comparison of rewards after 100 episodes for a trained agent, vs. the random agent. Means are shown in the violinplots as blue horizontal lines, and the mean of the trained agent is extended as the dashed purple line for comparison with that of the random agent.

After training, we test the agent in the environment and provide 5 example runs in the supplementary. An example of the state at the end of one run is shown in Fig. 2(b) and indicates an intriguing pattern where the defects are arranged at large distances from each other, resulting in a curved domain wall and considerable curl (and thus high reward). It should be noted that the defects have E-fields aligned along [10] and [0$\bar{1}$], thus the **P** vector would want to rotate such that it can ideally be along [1$\bar{1}$]. This is achieved for more than half of the lattice sites in Fig. 2(b), but not near the defect at the upper right. This is transient behavior likely caused by favorable initialization of polarization in the [$\bar{1}$0] direction, that seeded a small domain that is overall unstable, but nonetheless exerts considerable influence due to strong nearest neighbor coupling term.

When tested against the actions of a random agent for 100 episodes, it is seen that the distribution of rewards is shifted towards slightly higher values, as shown in the violinplot in Fig. 2(c). The tail of the distribution is also somewhat denser, indicating several episodes with considerably large rewards. This suggests that the agent has indeed learned a policy superior to random defect positioning. Note that this is not a simple task in a highly stochastic environment.

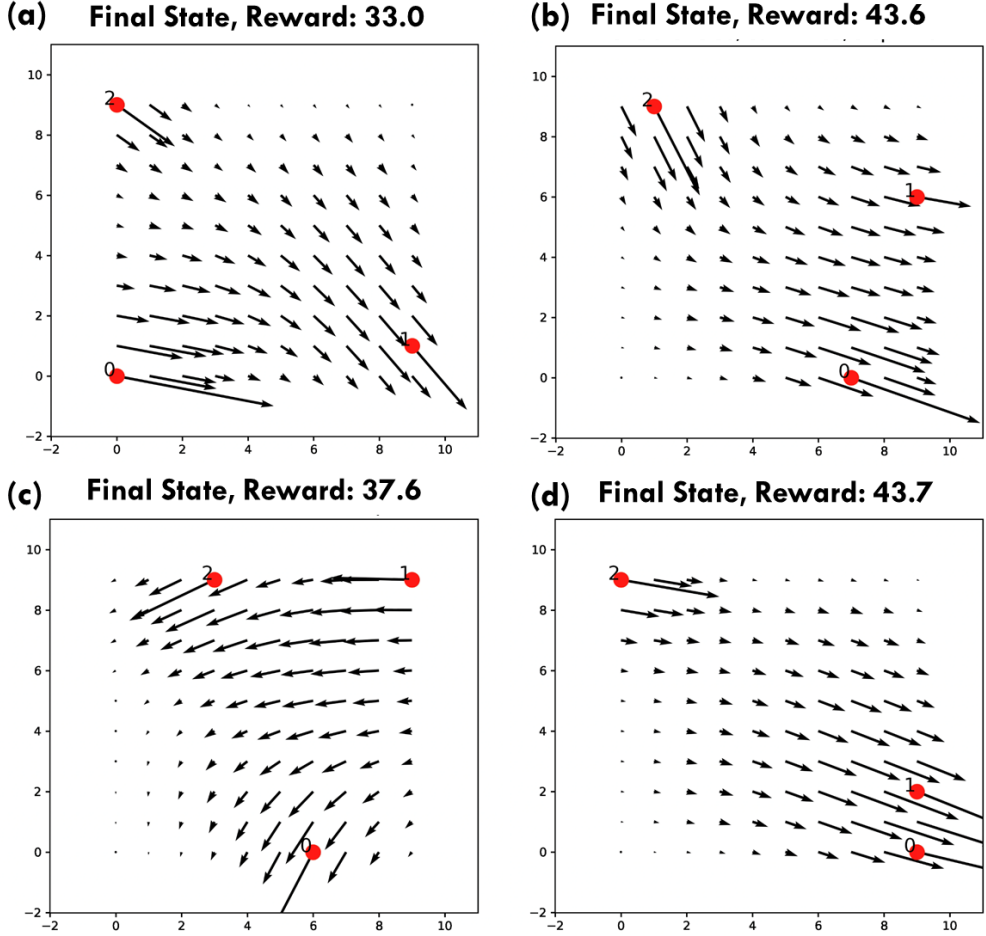

**Figure 3:** Final states and their respective rewards, for four runs of the (trained) RL agent are shown in (a)-(d). Triangular configurations appear to be preferred.

Four alternative endpoints are shown in **Figure 3**, and the initial seeding of the domains has severe consequences for the final polarization state in each episode. At the same time, three of the four final states have overall **P** vectors towards the energetically favored orientation, indicating some relaxation towards the thermodynamic minimum. Notably, the majority of runs end with the defects positioned in some type of triangular, but spatially distant, pattern which would seem to indicate the length scale for defect positioning for optimal polarization curl is quite large. One way to understand this is by considering the free energy functional equation (1), which can be written as $F_{Landau} + F_{coup} + F_{Elec}$. Given the choice of strong nearest neighbor coupling ($k = 22.5$), when compared to the Landau coefficients, the $F_{coup}$ and $F_{Elec}$ terms exert stronger influence. There will exist a considerable drive to rotate polarization quickly towards the direction of the defect-induced field, given that the field strengths are high. The strong coupling then drives nearest neighbors to quickly align with the polarization at the defect site. Away from the defect, the effects of this coupling gradually decay. The same phenomena can cause the 'kinetically stuck' configuration where the strong neighbor coupling prevents rapid reversal of the polarization from the unfavorable orientation.

**Comparison to simpler single-defect optimization**

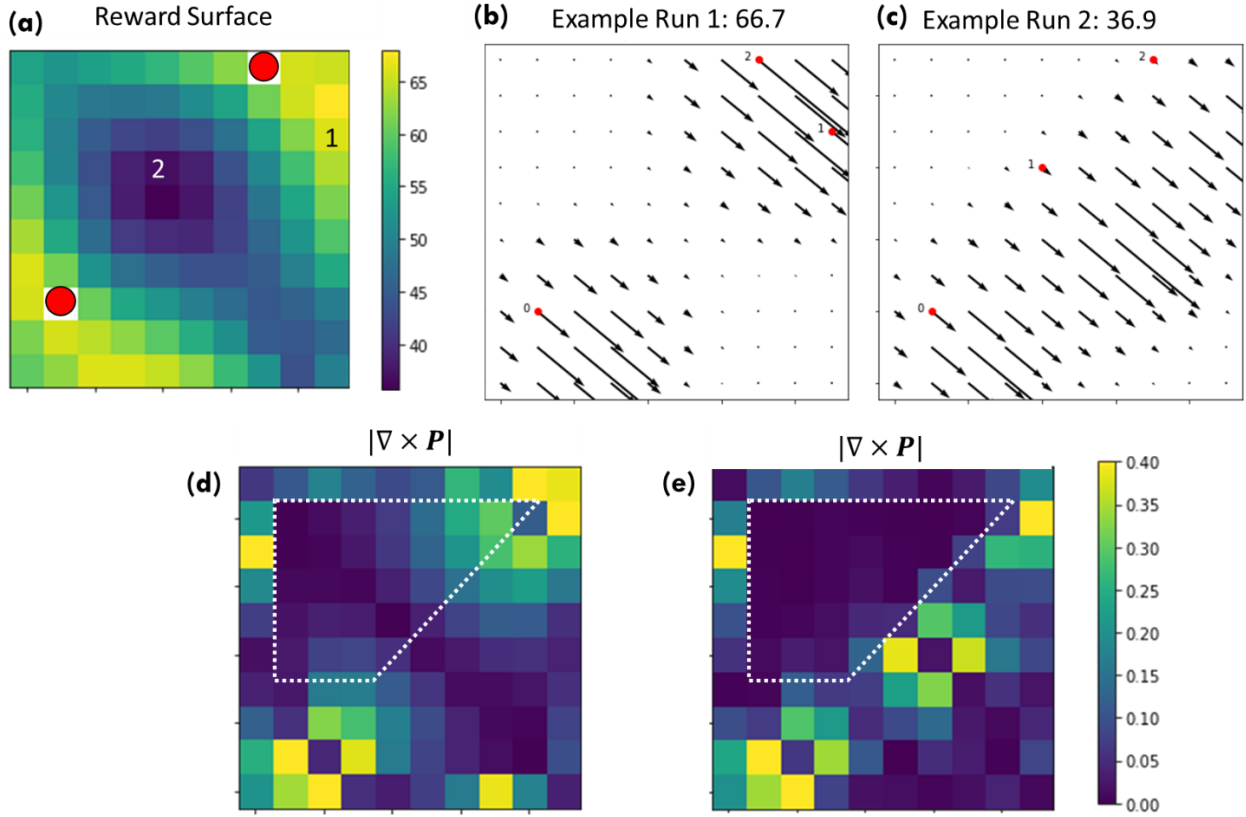

**Figure 4: (a)** Reward surface for moving a defect to any of the (unoccupied) lattice sites, given two fixed defects shown in red. (b) Simulation assuming the defect is placed in the site marked (1) in (a). (c) Simulation assuming the defect is placed in the area marked (2) in (a). (d) The curl magnitude map of the configurations in (b) and (c) are shown in (d) and (e), respectively.

To attempt to validate the learned policy, we performed a simple test where we fixed two defects near opposite corners, and then investigated the reward surface for positioning the remaining defect at all the other lattice sites. That is, the two red defects in **Fig. 4**(a) are fixed, and we simulate the reward the agent would have received if a third defect is moved to any of the other unoccupied lattice positions. This simulation is conducted 10 times for each lattice site, for 500 time steps for each run, and the averaged reward surface is shown in Fig. 4(a). Interestingly, this plot shows the maximum reward is found with positioning reminiscent of the agent's learned policy, in that the three defects adopt a large-scale triangular placement. An example of the polarization profile with the defect placed in the most favorable position is shown in Fig. 4(b) (marked as (1) in Fig. 4(a)). The reward surface indicates that the regions to the lower right of the static (immobile) defects are where the next defect should be placed to maximize reward. Moving the defect to a more central location, as in the position marked (2) in Fig. 4(a), leads to a situation where the polarization becomes more uniform and is visualized in Fig. 4(c). To investigate the origin of this behavior, we plotted $|\nabla \times \boldsymbol{P}|$ for both situations, in Fig. 4(d) and Fig. 4(e) corresponding to the states in Fig.

4(b) and (c), respectively. Although moving the defect to the more central position does induce higher magnitudes of curl to the lower-right side of the defect, it also significantly reduces the magnitude of the curl to the upper left side. As a guide to the eye, the area enclosed by the dashed line polygon shows the considerably reduced curl in this region in Fig. 4(e) as compared to (d). As such, we also do not see many runs where the trained agent adopts a configuration such as the one in Fig. 4(c).

**Attempting energetically unfavorable configurations**

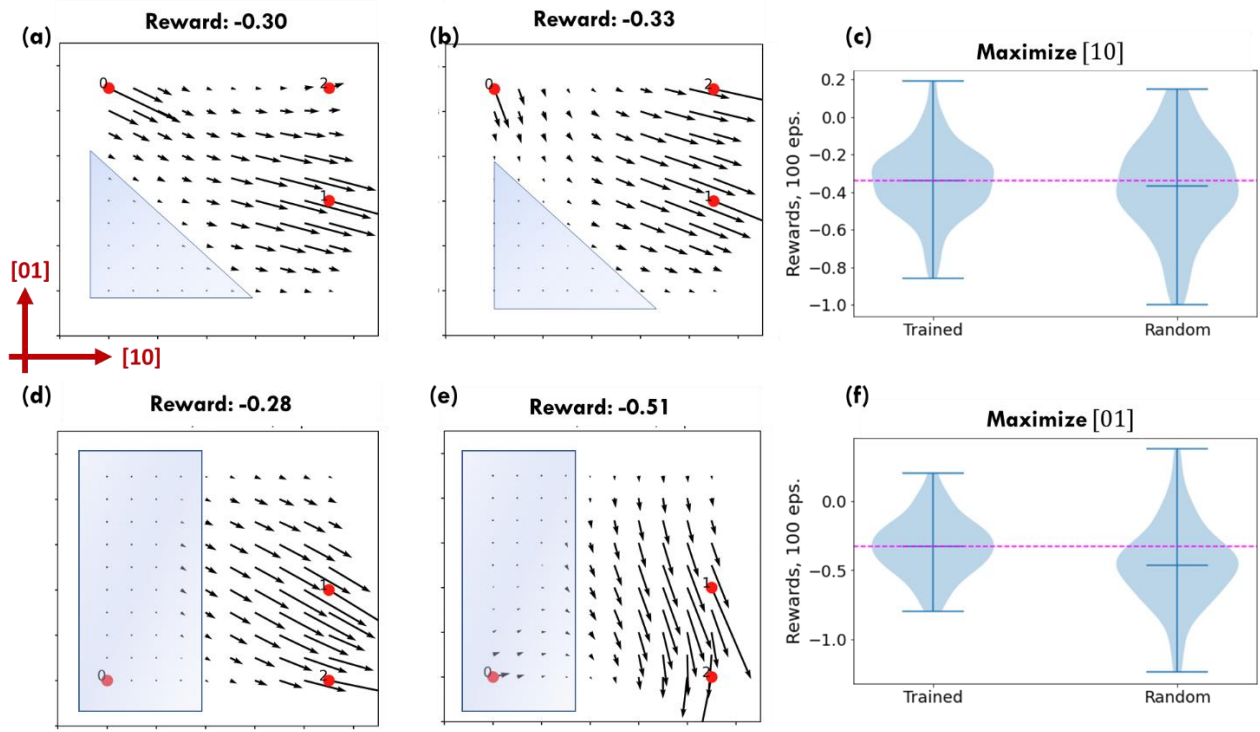

**Figure 5: Training agents to try to reach unfavorable states.** (a,b) Two final states of a trained agent attempting to maximize the polarization along [10] direction. (c) Comparison of rewards after 100 episodes, for the trained as opposed to random agent. (d,e) Two final states of a trained agent attempting to maximize the polarization along the [01] direction. (f) Comparison of rewards after 100 episodes, for the trained as opposed to random agent. In both violinplots in (c) and (f), the means are plotted as solid blue horizontal lines, and the mean of the trained agent is extended as a dashed purple line.

We are not limited to only maximizing the polarization curl. The environment can be modified to suit any reward. Here, we can for instance try to reward the agent for aligning the polarization along given directions, such as the [10] direction. This is calculated by simple dot product of the polarization with this target vector. Since there is no external applied field the default direction will be somewhere along [1$\bar{1}$], so aligning the P vector along [10] is an inherently energetically unfavorable situation, i.e. we would not expect positive rewards.

Nonetheless, it can be interesting to observe the strategies used to minimize the loss. Shown in Figure 5(a,b) are two example runs of the trained agent. The *x*-component is already favorably aligned, but the *y*-component of polarization attempts to account for the negative $E_y$ component of each defect and rotate down. Large rotations in this orientation will result in more negative rewards. To minimize this, one strategy that appears to be followed is to position defects in such a way so as to induce a substantial region with minimal polarization magnitude (shaded triangles in Fig. 5(a,b)). When compared to random movement of the defects, this strategy appears to be somewhat more successful, and incur less extreme losses, as shown in Fig. 5(c).

The same arguments apply to the case of aligning the polarization along [01]. Since this is impossible, the default method to minimize the loss appears to be to try to induce large regions of very small or zero polarization, as indicated by the shaded rectangular regions in Fig. 5(d,e). When compared to the random agent, this is substantially more helpful (Fig. 5(f)).

**Discussion/Summary**

This study illustrates how recent advancements in machine learning can be used to aid in microstructural design, by reformulating the problem as a reinforcement learning environment. It is expected that such a strategy will be beneficial for a variety of inverse materials design and optimization problems, although significant challenges remain with respect to underlying simulations being fast enough to enable RL agents to operate within them. In this respect, recent successes in utilizing deep neural networks for surrogate modeling, as well as fully differentiable programming frameworks that are GPU-accelerated can be expected to provide the necessary bridge.[20] From an experimental point of view, it is already possible to insert and create defects locally, e.g. with a scanning probe microscopy tip to inject oxygen vacancies.[21] Similarly, the electron beam in scanning transmission electron microscopy can be used to inject local charges, and subsequently read out polarization. Here, the polarization read-out has been demonstrated by multiple groups for over a decade,[22-26] and electron beam manipulation via e-beam has been demonstrated by e.g. Ferris[27] and Taheri.[28] From a purely simulations point of view, RL agents can provide significantly improved understanding of the underlying dynamics of the system. This is because RL policies are required, explicitly or implicitly, to be able to model state transitions and therefore future outcomes to current actions. Therefore, inspection of the derived strategies can be highly useful in determining features such as characteristic length scales and the effects of local fluctuations in the functional response.

**Acknowledgements**

The ferroelectrics simulations were supported by the US. DOE Office of Science, Basic Energy Sciences, Materials Sciences and Engineering Division. The Reinforcement Learning effort was supported by and conducted at the Center for Nanophase Materials Sciences, a US DOE Office of Science User Facility.

**Conflict of Interest**

The authors have no conflicts to disclose.

**Author Contributions**

RKV created the RL environment, optimized the RL agents, generated figures and co-wrote the paper. SVK analyzed data, assisted debugging, and co-wrote the paper. EO assisted with coding of the environment and hyperparameter optimization.

# Discovering mechanisms for materials microstructure optimization via reinforcement learning of a generative model

Rama K Vasudevan Erick Orozco, and Sergei V. Kalinin

**Supplementary Table S1**

This table lists the simulation parameters used in the manuscript.

| Simulation parameter | Value |
|---|---|
| $\alpha_1$ | -1.85 |
| $\alpha_2$ | 1.25 |
| $k$ | 22.5 |
| $\alpha_{dep}$ | 0.35 |
| $N$ | 10 |
| $\gamma$ | 1.0 |
| Defects E-field $(x,y)$* | $(15.05 + \varepsilon, -12.05 + \varepsilon)$ |

Note: Small noise ε is added to the E-field from the defects, on the order of 1% for added stochasticity

**Supplementary S2**

A full Jupyter notebook, which can be run on Google Colab, is available and which contains all the necessary code to reproduce every result in this paper.

Google Colab link

**Supplementary Videos**

Videos of trained agents are available and included with this manuscript for the agents trained. These can be downloaded via the same Colab link above. The videos titles and corresponding details are:

| Filename | Details |
|---|---|
| 'maxcurl_3def_run=k.mp4' (k=0-4) | Runs of trained agent from which the results in Figure 2-4 |
| 'MaxPOr_10_3def_run=k.mp4'(k=0-4) | Runs of trained agent shown in Figure 5(a-c) |
| 'MaxPOr_01_3def_run=k.mp4'(k=0-4) | Runs of trained agent shown in Figure 5(d-f) |